\documentclass[prb,twocolumn,aps,showpacs]{revtex4}
\usepackage{graphicx,color}
\usepackage{latexsym,amssymb}

\def\be{\begin{equation}}
\def\ee{\end{equation}}

\def\bc{\begin{center}}
\def\ec{\end{center}}
\def\ds{\displaystyle}
\def\vr{\vec{r}}
\def\vrp{\vec{r'}}

\begin{document}

\title{Absence of diffusion in certain random lattices: Numerical evidence.}

\author{P. Marko\v{s}}
\affiliation{%
Department   of Physics, FEI STU, 812\,99 Bratislava, Slovakia}

\begin{abstract}
Two numerical experiments are performed  to demonstrate the physical character of
electron localization in a  disordered
two-dimensional lattice. In  the first experiment we solve the time-dependent Schr\"odinger 
equation and show   that  the disorder prevents
electron  diffusion. Electron becomes spatially localized in a  specific area 
of the system.  The second experiment analyzes  
how an electron propagates through a disordered sample. 
 In strongly disordered systems,  we identify 
a narrow channel through which an   electron propagates from one side of the sample to the
opposite side.  We show that  this 
propagation is qualitatively different from the propagation of a classical particle. 
Our numerical analysis  confirms  that the electron localization
is a quantum effect caused  by the wave character of electron propagation
and has no analogy in classical mechanics.
\end{abstract}


\maketitle

\section{Introduction}

The electron localization in disordered systems  \cite{1}
is responsible for a broad variety of transport phenomena experimentally observed in
mesoscopic systems: the non-Ohmic behavior of  electron conductivity, weak
localization, universal conductance fluctuations, and strong electron localization.
\cite{MacKK,Janssen}

Localization arises  in systems with random potential. 
Let us consider the time evolution of a quantum particle located at a time $t=0$ 
in a certain small area  of the sample.  For $t>0$, 
 the electron wave function scatters
spatial inhomogeneities (spatial fluctuation of the potential).
Multiple reflected components of the wave function  interfere with each other.
As Anderson \cite{1} proved,
this interference can abolish  the  propagation. \cite{ziman,ATAF,WaveP,2}
As a  result, 
 wave function will be  non-zero only within a   specific area, determined by initial 
electron distribution, 
and decays exponentially as a function of the distance from the center of localization.
The probability to find an electron  in its initial  position
is non-zero for any time $t$, even when time increases to infinity, $t\to\infty$.

Similarly to the quantum bound state, the spatial extent of the 
localized wave function is finite.  
However, the physical origin of  the localization differs: 
a  bounded particle is trapped in the potential well,  
while localization
results from interference of  various 
components of the wave function scattered by randomly distributed fluctuations of the
potential. 

Localized electrons cannot conduct electric current. 
Consequently,  the probability of 
electron transmission, $T$,
through a disordered system decreases exponentially as a function of the system length $L$:
$T\propto \exp -L/\xi$. The length $\xi$ is called localization length.
Those materials that  do not conduct  electric current  due to electron  localization 
are called Anderson insulators.

In spite of significant theoretical effort, our understanding of electron localization is still
not complete.   Rigorous analytical results were obtained only in the limit of weak randomness,
where perturbation theories  are applicable. \cite{LSF,DMPK,PNato,Been}
In the localized regime,  we do not have any small parameter, so no perturbation analysis is possible. 
Also, as we will see below, the transmission 
of electrons is extremely sensitive to the change of sample  properties. 
In particular, small local change of random potential might cause the change of the transmission amplitude
in many orders of magnitude. Clearly, analytical description of such systems is difficult. 
Fortunately, it is rather easy to simulate the transport properties numerically. In fact,
many quantitative data  about the electron localization was  obtained numerically.
\cite{PS,McKK,KK,2}

In this paper we describe  two simple  numerical experiments which demonstrate  the  
key features  of quantum localization.
In Section \ref{model} we introduce 
the Anderson model that  represents the most simple model for study of the
electron propagation in the two-dimensional system with a random potential.
In Sections \ref{diffusion} and  \ref{localization}
we show 
how randomness influences the ability of an electron to propagate at large distances.
We solve numerically the time-dependent Schr\"odinger equation and 
confirm that after a certain time electron diffusion ceases.
The electron becomes spatially localized in certain part of the disordered lattice.
This  numerical experiment  reproduces the Anderson's original problem.\cite{1} 

In Section \ref{path}
simulate the  scattering experiment. We consider an electron approaching  the 
disordered system from outside, and
calculate the amplitude of the transmission through the sample.   
Since the transmission depends on the actual realization of random 
potential, we can, by a small local  change of the potential,
estimate the probability that electron propagates through any  given  sample area.
In this way, 
we investigate    the spatial distribution of the electron inside  disordered  
sample.
For weak disorder, we find that electron is homogeneously distributed throughout the sample.
On the other side, 
in the localized regime we show that electrons propagate through  narrow spatial channel 
across the sample.  
Although this channel resembles the trajectory of classical particles,
we argue that the electron still behaves a quantum particle.  We  demonstrate the
wave character of the electron propagation  
by a simple numerical experiment.

Both numerical experiments confirm the main feature of  electron localization: it 
has its origin in 
the  wave character of quantum particle propagation. There is no localization
phenomena in classical mechanics.

\section{The model}\label{model}

\begin{figure}
\bc
\includegraphics[width=7cm,clip]{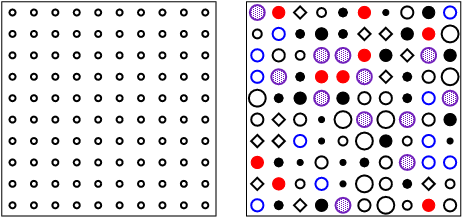}
\ec
\caption{
Left:  A regular  two-dimensional lattice is  a periodic arrangement of
identical atoms in a rectangular lattice.  Right:  A disordered lattice whose  sites are
randomly occupied by different atoms.
The closest distance between two  neighboring  atoms is $a$.}
\label{f-1}
\end{figure}

Left Figure  \ref{f-1} represents  the two-dimensional lattice created by regular 
arrangement of atoms. 
We consider one electron per atom and define  its local energy $\epsilon(\vr)=\epsilon_0$.
 If the electronic wave functions of  neighboring atoms overlap,
electrons can propagate through the lattice.
The periodicity  of the lattice creates  
a conductance band, $\epsilon_0-4V\le E\le \epsilon_0+4V$, \cite{economou}
where $V$   is given by  the overlap of electron wave functions
located in neighboring sites.

A disordered two-dimensional lattice is shown in right Fig. \ref{f-1}. 
Now, lattice  sites are occupied 
by different  atoms. Therefore, both the energy of  the electron  on a given
site,
$\epsilon(\vr)$,
and 
the hopping term between two neighboring atoms,  $V(\vr-\vrp)$, become position-dependent.
In our analysis we assume that energies $\varepsilon(\vr)$
are randomly distributed  according to  the Box 
probability distribution, $P(\epsilon) = 1/W$ if 
$-W/2\le \epsilon<W/2$, otherwise  
$P(\epsilon)=0$.
We also require these  random energies on different sites to be statistically independent and
assume that the hopping amplitude $V(\vr-\vrp)\equiv V$.
Although such random lattice is rather unrealistic, it imitates  all physical features of
a disordered electron system. 
The random energies  $\varepsilon(\vr)$ simulate  random potential.  

Thus, our random model is  characterized by two parameters: 
$W$ represents  the strength of the disorder and 
$V$ determines the hopping amplitude. 
Note that  $V$   defines the energy scale, so we  have 
only one parameter: the ratio $W/V$ that  we use as a measure of 
the strength of the disorder. 

Let us  assume that  at  a time $t=0$
a  quantum particle is located in the position  $\vr_0$. Initial wave function is
\be\label{dvax}
\Psi(\vr,t=0) = \delta(\vr-\vr_0).
\ee
We want to estimate the probability of the electron still being in its original position
in an infinite time $t\to\infty$.

The time evolution of the electron wave function 
is determined by the Schr\"odinger equation,
\be\label{ham}
i\hbar\ds{ \frac{\partial\Psi(\vr,t)}{\partial t}} 
=  \epsilon(\vr) \Psi(\vr,t) +
V\sum_{\vrp} \Psi(\vrp,t),
\ee
where $|\vr-\vrp|=a$ is the lattice constant.
Equation (\ref{ham}) defines  the Anderson model.

Let us take   the zero disorder case, $W=0$ first.  The electron located 
at time $t=0$ in a specific  lattice site, 
 will diffuse to the neighboring sites.  In the limit of
an infinite time $t\to\infty$, the electron will 
occupy  all sites of the lattice.
Consequently, the  probability to find it in its  original position 
equals zero (or, more accurately, it is approximately proportional to  $1/$(lattice volume)).  

In disordered lattice $W\ne 0$,   electron propagation  depends on
the strength of the disorder. 
Intuitively, one expects a very weak disorder not to affect the diffusion considerably, 
but  a  sufficiently strong
disorder should stop the  diffusion. 
Then there  should be a critical value  $W_c$: 
diffusion continues forever  when $W<W_c$
but ceases  when $W>W_c$.
In the original paper \cite{1} Anderson derived the equation for the critical disorder as
\be\label{ek}
\ds{\frac{W_c}{V}} = 2eK\ln (eK).
\ee

According to Eq. (\ref{ek}), the critical disorder depends only on the lattice connectivity  $K$
(the number of nearest neighbor sites). Nowadays, we know 
\cite{wegner,AALR} that
the dimension $d$ of the lattice is a more important parameter.
 In the absence of a magnetic field and of electron
spin, all states are localized in disordered systems with dimension $d\le d_c = 2$. Therefore,
the critical disorder $W_c=0$ for $d=2$ 
and is non-zero in systems with higher dimensionality $d>2$.

\section{Diffusion}\label{diffusion}

Now we demonstrate the Anderson's  ideas in numerical simulation.
We examine  how the electron diffuses in the
disordered lattice, defined by Eq.  (\ref{ham}).
The size of the system is $L\times L$, where $L=2048a$ for
weakly disordered samples and $L=1024a$ for systems with a stronger disorder ($W/V>4$).

\begin{figure}
\bc
\includegraphics[width=6.0cm,clip]{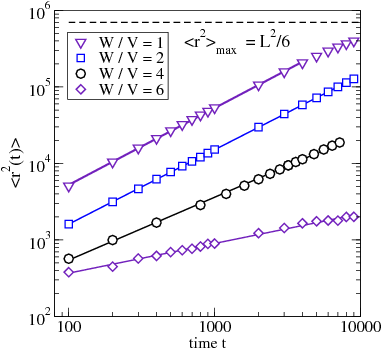}
\ec
\caption{(Color online) The quadratic displacement $\langle r^2(t)\rangle$ (in units $a^2$)
as a function of time $t$. Time is measured in $\hbar/V$.  The size of the system is 
$L\times L$ where $L=2048a$  ($L=1024a$ for $W/V=6$).
Note the logarithmic scale of both axes.
In  weak disorders, we expect the electron to diffuse, so
that  $\langle r^2(t)\rangle = 2D t$, in accordance with Eq. (\ref{dif}). 
Numerically, we  find
that $\langle r^2\rangle = 2Dt^\alpha$ with $\alpha = 1.004$ for  disorder $W/V=1$
 and $\alpha = 0.98$ for $W/V=2$. 
The corresponding diffusive constants are  $D=25.7$  and  $9.1$
(in units $a^2V/\hbar$). 
Only the data for time $t<4000 \hbar/V$ were used for $W/V=1$, since in  longer time
the  electron could  reach the edge  of the sample.
The dashed line represents  the limit $\langle r^2\rangle_{\rm max}=L^2/6$, 
given by Eq. (\ref{max}).
For stronger disorders, the  time evolution of the wave function is not
 diffusive. We find the  exponent $\alpha \approx 0.82$ ($W/V=4$)  and  $\alpha\approx 0.39$
($W/V=6$). 
}
\label{w4-d}
\end{figure}

\begin{figure}
\bc
\includegraphics[width=6.0cm,clip]{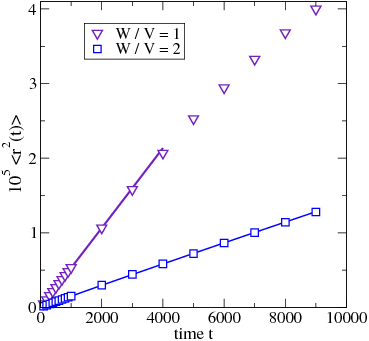}
\ec
\caption{(Color online) The same data as in Fig. \ref{w4-d}, but on a linear scale.
Only the data for small disorder is shown.
Note that for $W/V=1$,  $\langle r^2(t)\rangle$ is  linear only when   $t< 4000 \hbar/V$. 
This is because the electron  already reaches  the edge  of the sample.
}
\label{dva}
\end{figure}

\medskip

First,  we need to define the initial wave function $\Psi(\vr,t=0)$.  
A more suitable  candidate than the $\delta$-function (\ref{dvax})  
is any eigenfunction of 
the Hamiltonian defined on  small sub lattice 
(typically  the size of $24a\times 24a$) located in the center of the sample.\cite{ohtsuki}
Usually we chose the eigenfunction which corresponds to  the eigenenergy 
closest to $E=0$ (the middle of the conductance band).

To see how  the initial wave function develops in time $t>0$,  we solve 
the Schr\"odinger equation (\ref{ham}) numerically and find the time evolution of the 
wave function $\Psi(\vr,t)$.
The numerical program is based on the alternating-direction 
implicit method  \cite{NRCPS,elipt}
used for the solution of elliptic partial differential equations. The algorithm is 
described in Appendix A. 

\begin{figure}[b]
\bc
\includegraphics[width=5.0cm,clip]{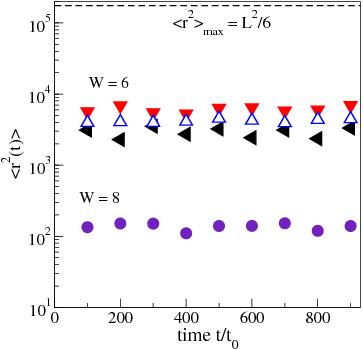}
\ec
\caption{(Color online) Quadratic displacement $\langle r^2(t)\rangle$ as a function of
time $t/t_0$, $t_0=1000\hbar/V$  for three systems of
the size  $L=1024 a$ and disorder  $W/V = 6$ (triangles).
Although $\langle r^2\rangle$ does not
increase when time increases, it
fluctuates as a function of time. The limiting value, $R^2$ (Eq. \ref{lim}) 
depends on the actual realization of the random disorder $\varepsilon(\vr)$ in the
 given sample.  the dashed line shows 
$\langle r^2\rangle_{\rm max}=L^2/6 =  174 762~a^2$ which is $50\times$ larger than actual  values
of $\langle r^2\rangle$.
For comparison, we also show  the quadratic displacement
for a system with stronger disorder, $W/V=8$, which is typically  $130 a^2$.
}
\label{w6-d}
\end{figure}

The ability of an electron to diffuse through the sample is measured by a  quadratic
displacement,  defined as
\be
\langle r^2(t)\rangle = \int d\vr r^2 |\Psi(\vr,t)|^2.
\ee
Figures \ref{w4-d}  and \ref{dva} show that in  weak disorders, $W/V=1$ and 2, 
 $\langle r^2(t)\rangle$ is a linear function of time $t$,
\be\label{dif}
\langle r^2(t)\rangle = 2Dt.
\ee
The parameter $D$ is a diffusive constant which  enters the Einstein formula for
electric conductivity $\sigma$,
\be
\sigma = e^2 D\rho.
\ee
Here $e$ is the  electron charge and $\rho$ is the density of states.\cite{2}

Since we analyze  only a lattice of a finite size, we  have to take into account that 
the $t$-dependence of the electron wave function 
might  be affected by the finiteness of our sample. In this case,  we not
 only observe the diffusion, but also the reflection
of the electron from  the edges.
Quantitatively, diffusion (\ref{dif}) is observable only when 
\be\label{max}
\langle r^2(t)\langle\ll
\langle r^2\rangle_{\rm max}=
\frac{1}{L^2}\int_{-L/2}^{L/2}
\int_{-L/2}^{L/2}\left(x^2+y^2\right)~dxdy
=\frac{L^2}{6},
\ee
where $\langle r^2\rangle_{\rm max}$
corresponds to the
 homogeneously distributed wave function, $|\Psi(\vr)|^2 = {\rm const} = 1/L^2$.

It might seem that the diffusion of electrons shown  in Figs. \ref{w4-d} and \ref{dva} contradicts 
the localization theory\cite{AALR} that  predicts all states to be localized in
two-dimensional systems. However, this is not the case. 
The prediction of the localization theory concerns   the limit of
an infinite system size. 
Physically, localization occurs  only when the size of the sample
exceeds the localization length, $L>\xi$. Since $\xi$ is very large in
weak disorder ($\xi\sim 10^6a$ when $W=1$),\cite{KK}  
we observe metallic behavior and diffusion of electrons
in Fig. \ref{w4-d}. 
Of course, even in the case of $W/V=1$  we would 
observe localization if much larger systems are taken into account.\cite{2}
In general, we can  observe the localization if we  either increase the size of the system
or reduce the localization length. The latter is easier to perform,
as  it requires us only to increase the
disorder strength $W$. We will do it  in the next Section.

\section{Absence of diffusion - localization}\label{localization}

\begin{figure}
\bc
\includegraphics[width=12.0cm,clip]{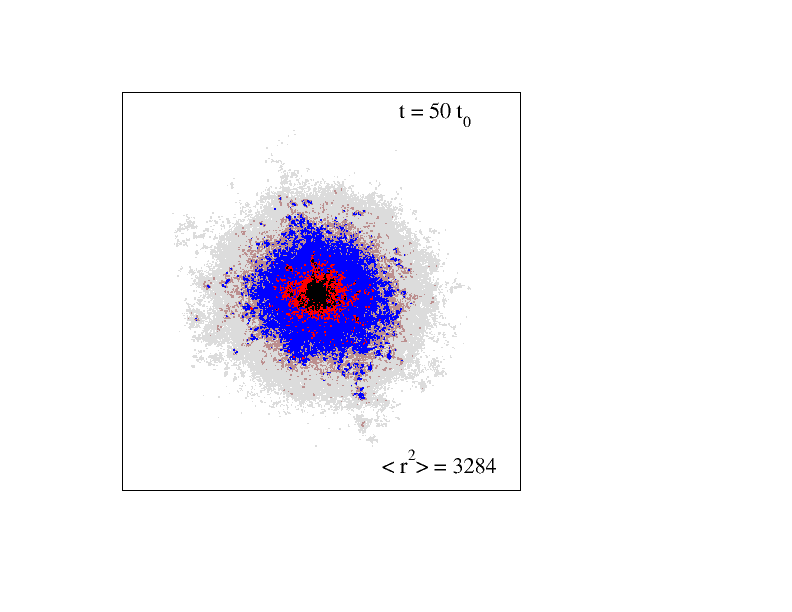}
\vspace*{-1cm}
\includegraphics[width=12.0cm,clip]{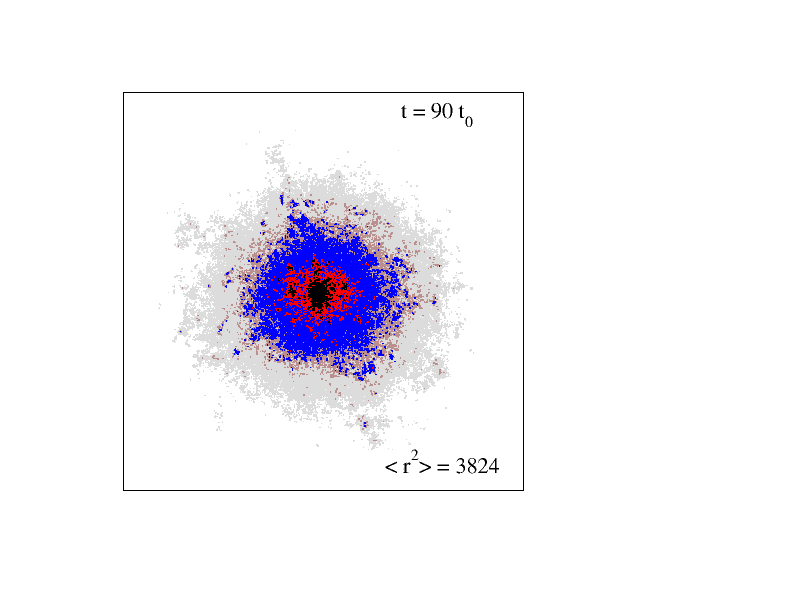}
\ec
\vspace*{-1cm}
\caption{(Color online) Spatial  distribution of an electron in sample with
disorder $W/V=6$.
The size of the lattice is $1024 a\times 1024 a$.
Time is given  in units
of $t_0 = 1000 \hbar/V$. The different colors show sites where 
$|\Psi(r)| >10^{-4}$ (gray), $> 5\times 10^{-4}$ (brown), $10^{-3}$ (blue),
$5\times 10^{-3}$ (red),
and  $> 5\times 10^{-3}$ (black). The probability to find an electron on any 
other site is less than $10^{-8}$.
}
\label{w6-t}
\end{figure}
\begin{figure}
\bc
\includegraphics[width=12.0cm,clip]{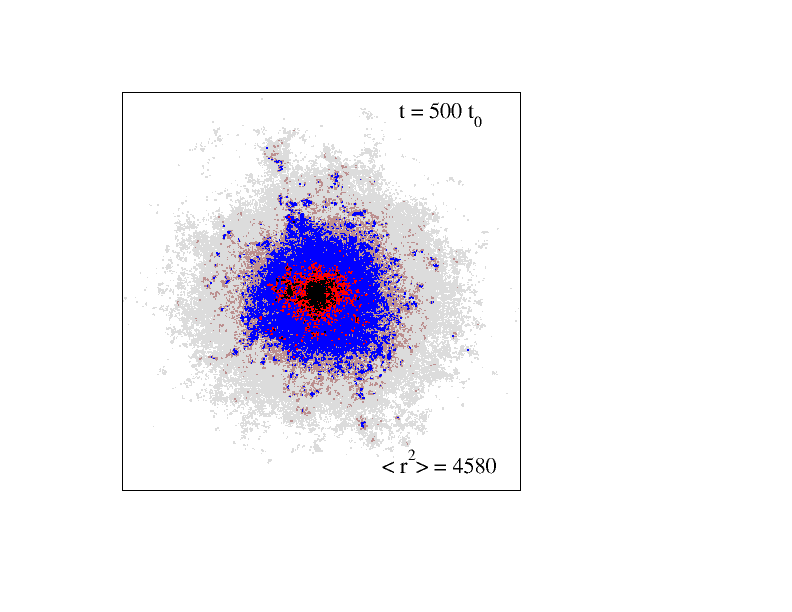}
\vspace*{-1cm}
\includegraphics[width=12.0cm,clip]{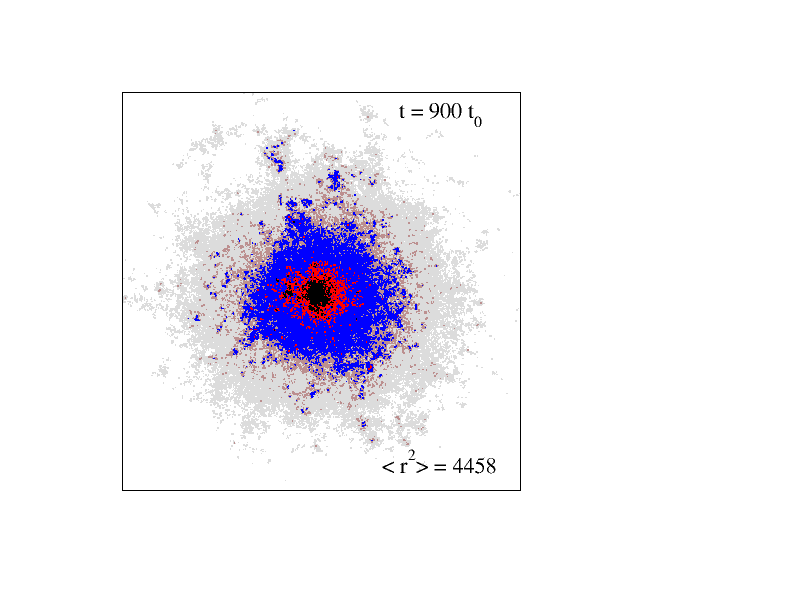}
\ec
\vspace*{-1cm}
\caption{(Color online) The same as in Fig. \ref{w6-t} only the  time is
$t= 500 t_0$ and $900 t_0$ ($t_0=1000\hbar/V$). 
}
\label{w6-tt}
\end{figure}

The data in Fig. \ref{w4-d} also  confirms  that the  time evolution of the wave function is
not diffusive when the disorder $W$ increases.   
Linear increase of $\langle r^2(t)\rangle$ is observable only for  short initial time interval.
For any longer time, 
the spatial extent of the electron  increases very slowly
and finally ceases (Fig. \ref{w6-d}). 
the electron becomes  localized.

To demonstrate the electron localization more explicitly,
we repeat  the experiment  in
Section \ref{diffusion} with a stronger disorder $W/V=6$. 
Similarly to the previous experiment, 
the initial wave function is non-zero in the small area 
$24a\times 24a$ located at the center of the sample. For shorter  times, we observe that
the spatial extent of the wave function increases. Then, after a while,
$\langle r^2(t)\rangle$ saturates:
\be\label{lim}
\lim_{t\to\infty} \langle r^2(t)\rangle =  R^2 \ll \langle r^2\rangle_{\rm max}.
\ee
Although the spatial distribution  of the electron varies in time,
$\langle r^2(t)\rangle$ does not longer increase
even if  the time $t$ increases ten and more times.

Figures  \ref{w6-t} and \ref{w6-tt} show the spatial distribution of the wave function,
$|\Psi(\vr,t)|$.  they represent the   lattice sites with 
$|\Psi(\vr)|>10^{-4}$. This means that  the probability to find the electron in any other 
lattice site is less than $10^{-8}$.

\begin{figure}
\bc
\includegraphics[width=11.0cm,clip]{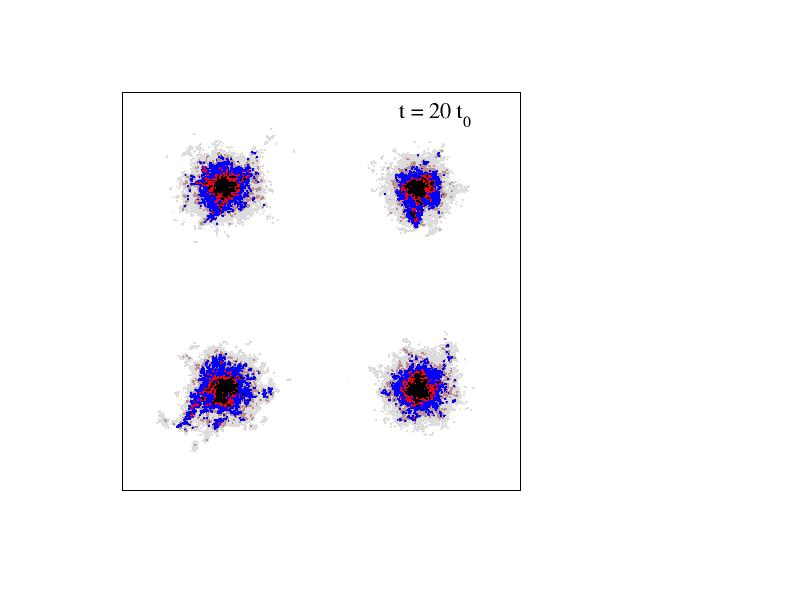}
\vspace*{-2cm}
\includegraphics[width=11.0cm,clip]{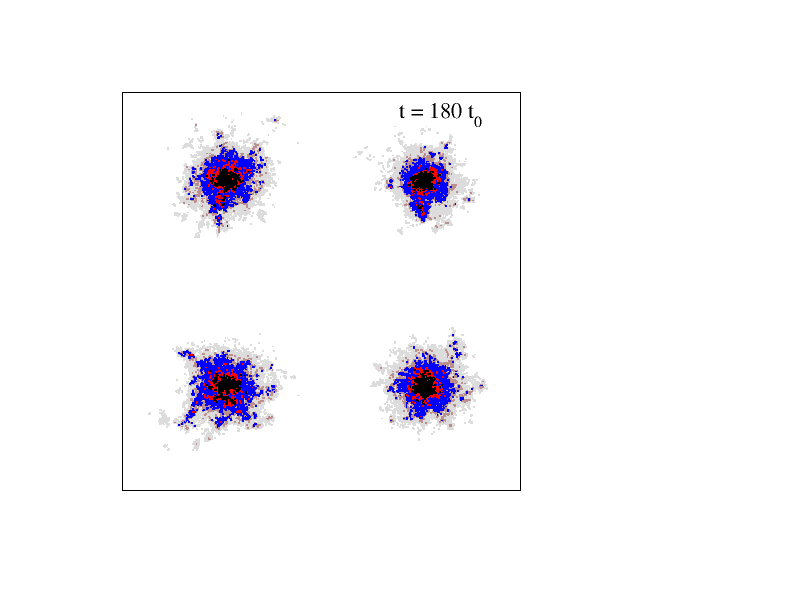}
\ec
\caption{(Color online) The time development of four electrons
located in  time $t=0$ in four different areas  of the same lattice.
The electrons do not leave the initial areas.
The size of the sample is $L=1024a$. Disorder $W/V=8$.
Again, time is measured in units of $t_0=1000\hbar/V$.
}
\label{W84-20}
\end{figure}

Note  that there is no potential well in the center of the sample where the electron is localized.
The only reason for the  electron being  localized in   the lattice center 
is the initial wave function, $\Psi(\vr,t=0)$, which was non-zero only in the center of the 
lattice. 
Applying   the initial wave function localized in  any
other area of the sample, we would  achieve    electron localization  in that
area. 
This is demonstrated  in Fig. \ref{W84-20}  showing   the time development of 
the wave functions of four electrons
in the same lattice. The initial position of the electrons is 
centered around four  points 
\be
x_\pm = L/2\pm L/4,~~~~y_\pm = L/2\pm L/4.
\ee
We see that in time $t>0$ each electron is localized around its initial position. 
This proves that localization is indeed the  result of
interference of wave functions. The electron is not trapped 
in  any potential well.  The localized state is not a bound state.

Figure \ref{W84-20} also shows that
the  localized states  are very sensitive to the realization of
the random potential.  
The spatial distribution of each electron reflects the local distribution
of random energies $\varepsilon(\vr)$.
This  is shown quantitatively
in Fig. \ref{w6-d} where we plot  $\langle r^2(t)\rangle$ as a function of time
for three 
different realizations of the  random disorder.  We see that  although all  three 
samples have  the same macroscopic parameter $W/V=6$, 
the limiting value   $R^2 = \lim_{t\to\infty}\langle r^2(t)\rangle$ is not universal but
depends on the actual
distribution of  random energies in the  given sample.
Moreover, $\langle r^2(t)\rangle$ fluctuates as a function of time $t$.

\section{Transmission through disordered sample:
How an electron propagates through disordered system?}\label{path}

\begin{figure}
\bc
\includegraphics[width=5.0cm,clip]{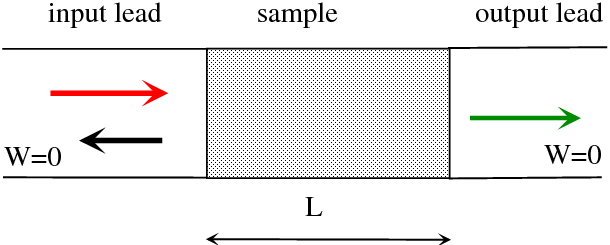}
\ec
\caption{Schematic description of the scattering experiment for measurement 
of the transmission. The sample is connected to two semi-infinite leads
represented by regular lattice  with zero disorder. 
Inside the sample,
the disorder is non-zero. If electron comes from the left, it either propagates through the sample
and contributes to the transmission, or  is reflected back to the left lead. 
}
\label{obr}
\end{figure}

Consider now another experiment, frequently used in the 
mesoscopic physics: We take a  disordered sample, the same as used in 
previous Sections, to examine  what  is the 
probability that an  electron propagates from one
side of the sample to the opposite side. 
Both in experiments and in numerical simulations the sample is connected to 
two semi-infinite, disorder-free leads which guide 
the electron propagation towards and out of  the sample
(Fig. \ref{obr}). An incoming electron either propagates through
the sample or is reflected back. The probability of  transmission, $T$, determines the
conductance, \cite{SE,landauer}
\be\label{se}
g = \displaystyle{\frac{e^2}{h}}~T.
\ee
Eq.  (\ref{se}) is commonly referred to as the Landauer formula. It  was originally derived
for a one-dimensional system  but can  also be  used for the analysis of
two- and more-dimensional samples. 
Since the width of the leads is non-zero the transmission $T$ can be larger than 1.
\cite{comment}
The transmission is calculated by the transfer matrix method described in Refs.
\cite{Ando-91,PMcKR,2}.

\begin{figure}
\bc
\includegraphics[width=5.0cm,clip,angle=-90]{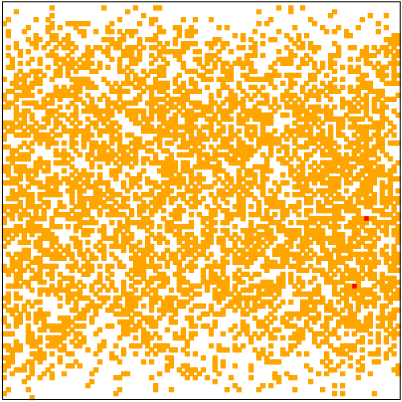}\\
~~\\
\includegraphics[width=5.0cm,clip,angle=-90]{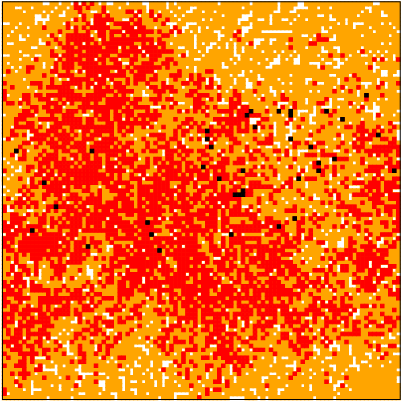}\\
~~\\
\includegraphics[width=5.0cm,clip,angle=-90]{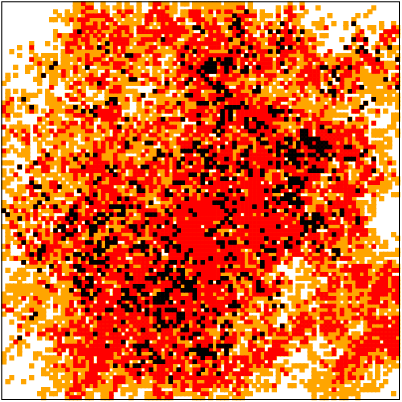}
\ec
\caption{(Color online)
Sensitivity of the transmission trough the disordered system to the change of
the sign of a single random energy $\vr_0$.
A change of the sign of the random energy on orange, red and black sites 
causes the change of the conductance by more than  1\%, 10\% and 100\%, respectively.
The transmission $T$  is  $4.998$, $0.52$  and 0.00084 for
the disorder $W/V=2$, 4 and  6 (from top to bottom).
The size of the system is $100a \times 100a$,
and the electron propagates from the left side of the sample to the  right side.
}
\label{w-L100}
\end{figure}

\begin{figure}
\bc
\includegraphics[width=5.0cm,clip,angle=-90]{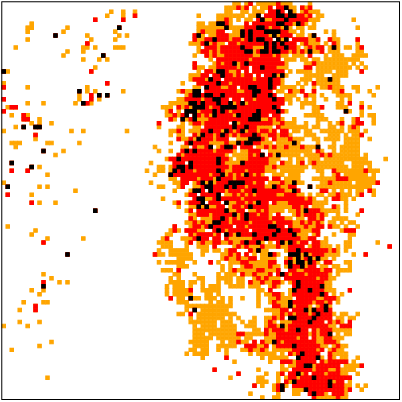}
\ec
\caption{(Color online)
The same as in Fig. \ref{w-L100}  but disorder $W/V=10$.
The transmission $T=9\times 10^{-15}$.
}
\label{w10-L100}
\end{figure}

Contrary to the diffusion problem, discussed in Sects. \ref{diffusion},\ref{localization},
in the present  experiment we do not analyze the time development of the  electron wave function.
Instead, we chose the energy $E$ of the electron ($E=0$, that is   the center of the
energy band), and calculate the time independent current transmission $T$  from the left
side of the sample to the right side. 

To show how electrons  are distributed within  the sample,
we apply Pichard's idea. 
\cite{PNato}
Let us 
change the sign of a single  random energy $\epsilon(\vr_0)$ at a site $\vr_0$:
$\epsilon(\vr_0)\to -\epsilon(\vr_0)$
and  calculate  how  this  change will 
influence the total transmission  $T$ of an electron through the sample.
We expect that $T$ is sensitive to the change of $\epsilon(\vr_0)$ only if
the electron occupies  the site $\vr_0$, i.e. when 
$|\Psi(\vr_0)|$ is large. Contrary,
if $|\Psi(\vr_0)|$ is negligible, then the change of $\epsilon(\vr_0)$ cannot affect
the transmission $T$. 

Thus, by the  comparison of the   transmissions through two systems  differing only
in the sign of the random energy $\varepsilon(\vr_0)$,
we can estimate whether the electron, propagating through the sample, travels
through the  site $\vr_0$ or not.  In repeating this analysis for   all lattice sites, we can
visualize the path of the electron through the sample.
For  numerical reasons,
we restricted the sample size to  $100 a\times 100 a$.

\begin{figure}
\bc
\includegraphics[width=10.0cm,clip,angle=-90]{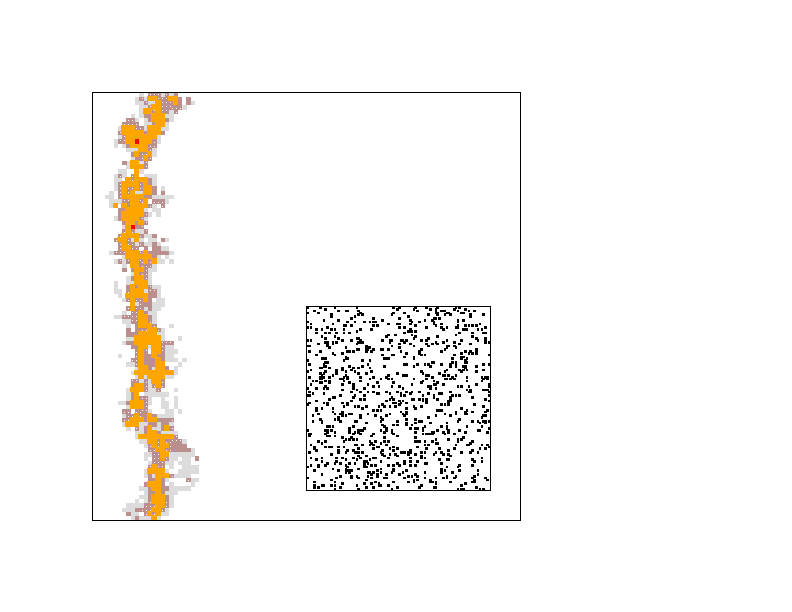}
\ec
\vspace*{-3cm}
\caption{(Color online) The same as in Fig. \ref{w-L100} but for  disorder
$W/V = 20$. The transmission is really small, $\ln T = -96$.  
The change of the sign of the random energy on 
gray, brown, orange and  red  sites 
causes a change of the logarithm of the 
transmission by more than  0.01\%, 0.1\%, 1\% and  10\%  respectively.
 Although it seems that  the path through the sample
is  determined by a valley in the potential landscape, this  is not the case.
The inset shows sites of the sample where the random energy  $|\epsilon| < 1$.
}
\label{w20-L100}
\end{figure}

\begin{figure}
\bc
\includegraphics[width=4.0cm,clip,angle=-90]{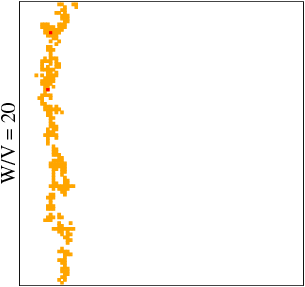}
~~~\includegraphics[width=4.0cm,clip,angle=-90]{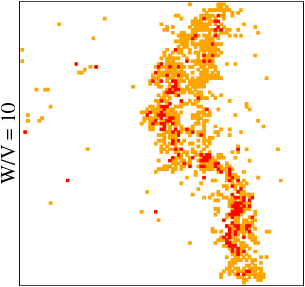}
\ec
\caption{(Color online)
The electron path through two strongly disordered samples: both samples have the same
realization of random energies. They differ only in the amplitude of fluctuations.
Shown are the lattice sites where the change of the sign of random energy causes the 
change of the logarithm of the transmission  by 1\% (orange) and 10\% (red).
We see that the electron prefers 
completely different trajectories through these samples. 
}
\label{w10-w20}
\end{figure}

Our results  are summarized in  Fig. \ref{w-L100}.
In  weak disorders, $W/V=2$, we see that 
the changing of only one random energy has an almost negligible influence on the transmission.
Typically, $T$ changes only by 1\% (or even less) when the sign of $\varepsilon(\vr)$ changes.
Also, all lattice sites  are more or less equivalent. 
We conclude that
in the course of the transmission the electron is ``everywhere'': it  propagates
through the entire sample as 
 a quantum wave. This observation is the key idea of the 
Dorokhov-Mello-Pereyra-Kumar theory of the electron transport in weakly disordered systems 
\cite{DMPK} and of the random matrix theory
of diffusive transport. \cite{PNato}

The homogeneity of the electron distribution
 gets lost when the disorder increases.\cite{muttalib,MMW}  The change of the random energy 
sign on some sites influences the transmission  more than  the same change on 
other sites. Some areas  of the sample seem not to be visited at all.
We can see the formation of the electron ``path'' through the sample.\cite{prior}
This path is clearly visible for very strong disorder  shown in Figs. \ref{w10-L100},
\ref{w20-L100} and \ref{w10-w20}.

However, we want to stress
that  even in  case of strong disorders
we cannot speak about the path in its  classical sense. Even if the electron path is well 
visible, 
 there  are still   other sites, often located on the opposite side of the sample,
that  influence the transmission as strongly as the sites on the main trajectory
 (Fig. \ref{w10-L100}).
This indicates  that the electron still feels the entire  sample and  its propagation is 
highly sensitive to any change of the realization of the random potential.

the  resulting trajectory cannot be identified with any valley or equipotential line  in the 
random potential landscape. To demonstrate this, we show in Fig. \ref{w20-L100} the trajectory
of an electron through an  extremely strongly disordered system 
($W/V=20$ - in this case, we consider the
change of the logarithm of the conductance). Although the trajectory
of an electron seems to be well defined,  there 
is no continuous potential valley  which might support the propagation. 
The  inset of
Thus, the choice of the
transmission path  is the  result of quantum interference: an electron arising  from the left  
inspects the  entire sample and finds the most convenient spatial 
``channel'' for its  propagation. We cannot 
speak about a trajectory in the sense of classical particles.

To support our  last claim,  let us  consider two samples, with 
the same realization of random energies $\varepsilon(\vr)$, but 
different    amplitudes of random energies: $W/V=10$ for the sample I
and $W/V = 20$ for the sample II: 
\be
\varepsilon(\vr)^{II} = 2\varepsilon(\vr)^{I}
\ee
for each site $\vr$.  With the help of  the above-mentioned method, we find 
the trajectories of electrons through these two  samples.
For the
propagation of classical particle both  trajectories (for the sample I and sample II)
should coincide. However, 
an  electron is not a classical particle.
As shown in Fig. \ref{w10-w20}, the ways how an  electron propagates through the 
two samples, I and II,  are completely different.
An increase of fluctuations of the random potential causes the electron to choose
a completely different route.

\section{Conclusion}

We  discussed two  features of localization of quantum particle in a
disordered sample. Firstly, we demonstrated numerically that the diffusion of the
quantum particle through randomly fluctuating potential ceases after certain time.
 The particle becomes spatially localized. 
The physical origin of  localization is different from the bounding of a particle in
a potential well. Localization is caused by a multiple scattering of the wave function
on randomly distributed impurities (fluctuations of the random potential).
It is  not due to the trapping of the particle in the potential well.

In the second part of the paper, we examined  the propagation of a quantum particle through
a disordered sample and discussed  how this propagation depends on the disorder. Again, 
we confirmed indirectly the wave
character of the propagation. 
This drove us to the conclusion that the  electron localization 
is a purely quantum effect without any analogy in classical mechanics.

In both numerical experiments, the key  condition for the localization to happen 
is the quantum coherence of the wave function. 
This is generally not fulfilled in  experiment, where the 
incoherent scattering - for instance the scattering of
electrons with phonons - plays the crucial role. 
As any incoherent scattering destroys quantum coherence,  the observing 
of electron localization experimentally requires  
the mean free path of incoherent scattering to be larger 
or at least comparable to size
of the sample. This happens at a very low temperature.
Of course, localization does affect the transport of electrons also at higher
temperatures.  These effects  are, however,  above the scope of  present discussion.

With  localization  being 
a wave phenomenon, we can expect  the similarity 
of quantum propagation with classical wave phenomena. \cite{dragoman} 
That enables  us  to observe localization in many other instances.
In particular,  we can expect that 
classical waves  - electromagnetic or acoustic -
will  also be  localized 
in a  disordered medium. \cite{S} The localization of microwave electromagnetic waves
was experimentally observed \cite{GG}. Another very interesting experiment\cite{qq} proves the 
weak localization of seismic waves.

\appendix

\section{Numerical solution of the Schr\"odinger equation}

To integrate the Schr\"odinger equation (\ref{ham}) numerically, we  first 
have to discretize the time derivative:
\be\label{der}
\displaystyle{\frac{\partial\Psi(\vr,t)}{\partial t}} = \frac{1}{\delta t}\left[\Psi(\vr,t+\delta t) - \Psi(\vr,t)\right],
\ee
where $\delta t$ is the time iteration step.
In order to define the iteration procedure for the 
calculation of  the wave function in $\Psi(\vr,t+\delta)$ in terms of $\Psi(\vr,t)$,
we also have  to determine the time in 
which the wave function on the right-hand side of Eq. (\ref{ham})
is calculated.  
The explicit method takes  the entire r.h.s. of Eq. (\ref{ham})  in  
time $t$. The  resulting iteration scheme  is simple,  the  time step $\delta t$, however,
has to  be very small  
to avoid  numerical instabilities.\cite{NRCPS} Other, more sophisticated, 
explicit methods based on the
Suzuki-Trotter formula  are described  in Ref.\cite{ohtsuki} In this paper, we use
the alternating-direction implicit iteration schema. \cite{NRCPS,elipt}
This method presses the  numerical integration of Schr\"odinger equation (\ref{ham}) 
in  two steps. Firstly,   we write
the Schr\"odinger equation (\ref{ham}) in its discrete form ($\vr = (x,y)$)
\be\label{ss-1}\begin{array}{ll}
\Psi(x,y,t+\delta t) &=  \Psi(x,y) +\displaystyle{\frac{V}{i\hbar}}\delta t
\Big[ 
\frac{W}{V}\varepsilon(x,y)\Psi(x,y,t)\\
& \\
& +\Psi(x+a,y,t+\delta t) + \Psi(x-a,y,t+\delta t)\\
& \\
& + \Psi(x,y+a,t) + \Psi(x,y-a,t) \Big]
\end{array}
\ee
Note that the wave function along the $x$ direction is  in time $t+\delta t$,
while the wave function in $y$ direction is  in time $t$.
The second step is to  put $t\to t+\delta t$  and re-write the iteration equation 
in the form
\be\label{ss-2}\begin{array}{ll}
\Psi(x,y,t+\delta t) &=  \Psi(x,y) +\displaystyle{\frac{V}{i\hbar}}\delta t 
\Big[ 
\frac{W}{V}\varepsilon(x,y)\Psi(x,y,t)\\
& \\
& +\Psi(x+a,y,t) + \Psi(x-a,y,t)\\
& \\
& +  \Psi(x,y+a,t+\delta t) + \Psi(x,y-a,t+\delta t) \Big].
\end{array}
\ee
Now the wave function along the $y$ direction is given  in time $t+\delta t$
and the wave function along  $x$ direction is in time $t$.

The advantage, that this algorithm presents is its  numerical stability,
 even for rather large values of  $\delta t$. 
The price for the numerical stability is that  we have to solve
$N$  ($N=L/a$) systems of $N$  linear equations in each iteration step. 
Fortunately, iteration schema (\ref{ss-1},\ref{ss-2}) only requires solution of 
the  three-diagonal system of linear equations, what is easy to calculate. 

Other modifications of the above  algorithm  are possible 
since we are free to substitute the wave function $\Psi(x,y,t)$ on the r.h.s of 
Eqs. (\ref{ss-1},\ref{ss-2}) by $\Psi(x,y,t+\delta t)$, or, eventually, by
$[\Psi(x,y,t)+\Psi(x,y,t+\delta t)]/2$.

To measure the accuracy of the numerical solution, we also calculate in each time $t$
the norm of the wave function,
\be
{\cal N} = \int d\vr |\Psi(\vr,t)|^2.
\ee
We obtained that for $\delta t=0.1\hbar/V$ the norm ${\cal N}$
is always very close to 1:
\be
|{\cal N}-1|<2\times 10^{-3}.
\ee

\bigskip

This work was supported by project APVV n. 51-003505  and project VEGA  2/6069/26.


\begin{thebibliography}{99}
\bibitem{comment}
For non-zero width of leads, the wave function in leads is a superposition of different modes.
Each mode is characterized by its transverse momentum, $k_{\perp\alpha}$. 
Since the electron is a quantum particle, $k_\perp$ is quantized. 
For each transversely wave vector, we have a longitudinal component $k_{\parallel\alpha}$
given by the dispersion relation  $E=E(k_{\perp\alpha},k_{\parallel\alpha})$. 
Note that  some values of $k_\parallel$ might be imaginary.
these modes $\alpha$ are called evanescent.
They do not transfer the current in leads, contrary to
propagating modes with real  $k_\parallel$. 
If we have  $N$ propagating  modes, 
then $T=\sum_{\alpha}^N\sum_{\beta}^N T_{\alpha\beta}$, where 
 $T_{\alpha\beta}$ is the probability that an electron coming from the left with 
a  wave vector $k_\alpha$ appears 
on the opposite side of the sample  with the wave vector $k_\beta$.
Since $\sum_\beta T_{\alpha\beta}\le 1$, the maximum   of $T$ is $N$.
\end{thebibliography}
\end{document}